\documentclass[a4paper,11pt]{article}
\pdfoutput=1 
\usepackage{jheppub} 
\usepackage[T1]{fontenc} 

\usepackage{jheppub}
\usepackage{color}
\usepackage{graphicx}
\usepackage{wrapfig,enumerate,slashed}
\usepackage{jheppub}
\usepackage[utf8]{inputenc}

\usepackage{tabularx}
\usepackage{placeins} 

\title{\boldmath 
Minimal flavor-changing $Z'$ models and 
muon $g-2$ after the $R_{K^*}$ measurement }

\author[a]{Stefano Di Chiara,}
\author[b]{Andrew Fowlie,}
\author[a]{Sean Fraser,}
\author[a]{Carlo Marzo,}
\author[a]{Luca Marzola,}
\author[a]{Martti Raidal,}
\author[a]{Christian Spethmann}

\affiliation[a]{
National Institute of Chemical Physics and Biophysics, R\"{a}vala 10, Tallinn 10143, Estonia}
\affiliation[b]{
ARC Centre of Excellence for Particle Physics at the Tera-scale,
School of Physics 
and Astronomy, Monash University, Melbourne, Victoria 3800 Australia}

\emailAdd{stefano.dichiara@kbfi.ee}
\emailAdd{andrew.fowlie@monash.edu}
\emailAdd{sean.fraser@kbfi.ee}
\emailAdd{carlo.marzo@kbfi.ee}
\emailAdd{luca.marzola@cern.ch}
\emailAdd{martti.raidal@cern.ch}
\emailAdd{christian.spethmann@kbfi.ee}

\abstract{
There has been a steady interest in flavor anomalies and their global fits as ideal probes of new physics. If the anomalies are real, one promising explanation is a new $Z'$ gauge boson with a flavor-changing coupling to bottom and strange quarks and a flavor-conserving coupling to muons and, possibly, electrons.
We point out that direct production of such a $Z'$, emerging from the collision of $b$ and $s$ quarks, may offer a complementary window into these phenomena because collider searches already provide competitive constraints. On top of that, we analyze the same $Z'$ scenario in relation to another long-standing discrepancy between theory and experiment that concerns the anomalous magnetic moment of the muon. By scanning the allowed $Z'$ coupling strengths in the low-mass region, we assess the compatibility of the signals from LHCb with the $Z'$ searches in the high energy LHC data and the measurements of the anomalous magnetic moments of the involved leptons. We also argue that observations of the latter can break the degeneracy pattern in the Wilson coefficients $C_9$ and $C_{10}$ presented by LHCb data. The $Z'$ model we consider is compatible with the new measurement of $R_{K^*}$, therefore it can potentially account for the long-standing deviations observed in $B$-physics.
}

\begin{document} 
\maketitle
\flushbottom

\section{Introduction}

Processes involving flavor-changing neutral currents are sensitive probes of new physics. 
In the Standard Model (SM), transitions such as 
$b \rightarrow s \ell^+\ell^-$ are loop-suppressed, but new particles can contribute at tree-level. The possible impact of new particles on these processes is usually analyzed by integrating out the heavy degrees of freedom and working with the effective Hamiltonian. 
For $b \rightarrow s$ transitions we have,
\begin{equation}
\mathcal{H}_{\rm eff}=-\frac{4G_F}{\sqrt{2}} \frac{e^2}{16\pi^2} V_{tb}^{\phantom{*}} V_{ts}^*
\sum_i (C_i O_i + C_i' O_i') + H.c.\,,
\end{equation}
which is expressed in terms of the effective operators $O_i$, $O_i'$ and the Wilson coefficients 
$C_i$, $C_i'$.
Several anomalies with respect to SM predictions have been measured, typically
at the $2\sigma$ level, but with increasing statistical significance. One notable anomaly of lepton flavor universality
is the measured
ratio of branching fractions 
$R_K = \mathcal{B}(B^+ \to K^+ \mu^+\mu^-)/
\mathcal{B}(B^+ \to K^+ e^+e^-) 
\simeq 0.745$~\cite{Aaij:2014ora}, 
as well as the recent measurement of
$R_{K^*} = \mathcal{B}(B^0 \to K^{*0} \mu^+\mu^-)/
\mathcal{B}(B^0 \to K^{*0} e^+e^-)\sim 0.7$ by 
LHCb~\cite{LHCB},
which has already prompted several 
studies~\cite{Capdevila:2017bsm,Altmannshofer:2017yso,DAmico:2017mtc,Hiller:2017bzc,Geng:2017svp,Ciuchini:2017mik,1592397,1592392}. Angular observables have become a popular testing ground,
and recent updates~\cite{ATLAS:2017dlm,CMS:2017ivg} 
have confirmed previous measurements in the decay
$B^0 \rightarrow K^{*0} \mu^+\mu^-$.
Including other transitions and a large number of
observables, global fits to the anomalies 
seem to be converging
on preferred sets of Wilson 
coefficients~\cite{Altmannshofer:2017fio,Descotes-Genon:2013wba,Descotes-Genon:2015uva}. However, it is not yet entirely clear which of these 
sets could be responsible for the anomalies. 
For example, refinement of the hadronic uncertainty 
in these analyses is an ongoing theoretical issue.

The general features of a $Z'$ gauge boson~\cite{Langacker:2008yv}
make it a good choice to generate  
$b \rightarrow s \ell^+\ell^-$ transitions,
as underlined by many previous studies~\cite{Descotes-Genon:2014uoa,Crivellin:2015era,Buras:2012jb,Gauld:2013qba,Buras:2013qja,Gauld:2013qja,Buras:2013dea,
Allanach:2015gkd,Crivellin:2015mga,Crivellin:2015lwa,Belanger:2015nma,Megias:2017ove,Megias:2016bde,Megias:2017dzd,GarciaGarcia:2016nvr,Carmona:2015ena}.
In this work, we follow a complementary approach
and consider what else can be gained from colliders
and other low-energy experiments.
Specifically, we examine the sensitivity of
the Wilson coefficients to constraints from the
direct production of a suitable $Z'$ gauge boson that can also 
explain the discrepancy 
between the 
measured value of the muon magnetic moment and the SM prediction.

We will focus here on scenarios in which the supposed new physics contributions affect exclusively 
$C^\ell_9$ and $C^\ell_{10}$, for $\ell=\mu,e$.
These coefficients are among those
currently favored
by global fits to the anomalies~\cite{Altmannshofer:2017fio},
as well as by
a recent analysis~\cite{Altmannshofer:2017yso},
which examines 
compatibility with $R_K$ and the new measurement of $R_{K^*}$. The relevant operators are
\begin{equation}
O_9^\ell = (\bar{s} \gamma_\mu P_L b) (\bar{\ell} \gamma^\mu \ell) \quad\text{and}\quad O_{10}^\ell = (\bar{s} \gamma_\mu P_L b) (\bar{\ell} \gamma^\mu \gamma_5 \ell). 
\end{equation}
Another motivation for this choice is that although the measurements of $R_{K^*}$ and $R_K$ suggest there is new physics which discriminates between muons and electrons, 
the LHCb data currently exhibit an interesting
degeneracy in $C^\ell_9$ and $C^\ell_{10}$ based on their statistical 
pulls~\cite{Altmannshofer:2017yso}: we argue that such degeneracy can be broken by analyses of the magnetic moments of the involved leptons in scenarios where the considered $Z'$ boson is well below the TeV scale. 
In addition, our analysis will show that the LEP bounds strongly disfavor scenarios where the speculated $Z'$ boson couples purely to electrons, besides quarks.

After introducing the adopted framework and detailing our methodology in Section~\ref{sec:The model}, we present the results of our analysis in Section~\ref{sec:A first analysis}. Our conclusions are offered in Section~\ref{sec:Conclusions}.

\section{A simple $Z'$ model} 
\label{sec:The model}

We assume a general interaction of a $Z'$ boson with quarks
$\bar{s} b$ and leptons $\bar{\ell} \ell$ described by the Lagrangian
\begin{eqnarray}
\mathcal{L} &\supset& 
\frac{g_2}{2c_W}
 Z'_\alpha
\left\lbrace
\left[
\bar{s}\gamma^\alpha( g^q_L P_L +  g^q_R P_R )b 
+ h.c.
\right]
+ 
\bar{\ell}\gamma^\alpha( g^\ell_V + \gamma_5 g^\ell_A )\ell 
\right\rbrace,
\end{eqnarray}
where $\ell=\mu,e$. The companion interactions with
neutrinos and up-type quarks 
are allowed by the SM gauge symmetry and may offer interesting
features which, however, we do not pursue in this work.
The quark flavor-violating couplings $g^q_L$, $g^q_R$ 
and the lepton flavor-conserving couplings $g^\ell_V$, $g^\ell_A$  are 
normalized relative to $g_2/(2c_W)$ from the Standard Model for convenience, 
$c_W$ being the cosine of the Weinberg angle.

After integrating out the $Z'$ and performing tree-level matching, 
the Wilson coefficients bounded by the LHCb results are related to the $Z'$ couplings by
\begin{equation}
\frac{e^2}{16\pi^2}
V_{tb}^{\phantom{*}} V_{ts}^* \cdot
\Big\{ C^\ell_9,\, C^\ell_{10} \Big\} 
= \frac{M_Z^2}{2M_{Z’}^2} \cdot
\Big\{g^q_L g^\ell_V,\, g^q_L g^\ell_A \Big\}.
\end{equation}
The Wilson coefficient $C_9^\ell$ encapsulates the vectorial couplings between the $Z'$ and leptons, whereas $C_{10}^\ell$ 
contains the axial couplings. As we can see, despite its simplicity, the $Z'$ model at hand can potentially account for deviations from lepton flavor universality.

The $Z'$ contributions to the magnetic moment of a charged lepton $\ell=\mu,e$ are given by
\begin{equation}
\Delta_{g-2}^{\ell}=\frac{1}{12\pi^2}\left(
\frac{g_2\,g_V^\ell}{2c_W}\right)^2\frac{m_\ell^2}{M_{Z'}^2}\,,
\end{equation}
for $C_9$ scenarios, and by
\begin{equation}
\Delta_{g-2}^{\ell}=-\frac{5}{12\pi^2}\left(
\frac{g_2\,g_A^\ell}{2c_W}\right)^2\frac{m_\ell^2}{M_{Z'}^2}\,,
\end{equation}
for $C_{10}$ scenarios,
where $M_{Z'}$ is the mass of $Z'$.
In our analysis 
we will refer to the following values for the measured
discrepancies of the involved lepton magnetic moments: $\Delta_{g-2}^e = (-10.5\pm8.1)\times 10^{-13}$, $\Delta_{g-2}^\mu = (290 \pm 90)\times 10^{-11}$~\cite{Giudice:2012ms,Jegerlehner:2009ry}.

The parameters to be scanned over are 
$M_{Z'}$ and the set of couplings
\begin{eqnarray}
\text{$g^q_L$, $g^q_R$, $g^\ell_V$ and $g^\ell_A$,}
\end{eqnarray}
which are for instance subject to the LHC dilepton searches that provide upper bounds on the production cross section times branching ratio as a function of $M_{Z'}$, $\sigma_{\ell\ell} \equiv \sigma(pp \to Z')\cdot\mathcal{B}(Z'\to \ell^+\ell^-)$.
Using the interaction Lagrangian above with specified input values of the $Z'$ mass and couplings, we numerically simulate the dilepton searches~\cite{ATLAS:2016cyf} yielding the constraint
\begin{eqnarray}
\sigma_{\ell\ell}^{\mathrm{sim}}(M_{Z'}, g^q_L,g^q_R, g^\ell_V,g^\ell_A)
\; \le \; 
\sigma_{\ell\ell}^{\mathrm{exp}}(M_{Z'}),
\end{eqnarray}
where $\sigma_{\ell\ell}^{\mathrm{sim}}$ is the result of our simulation, after the usual kinematical cuts, and $\sigma_{\ell\ell}^{\mathrm{exp}}$ is the experimental upper bound.

The most stringent limits on the couplings of the $Z'$ to electrons
originate from LEP electroweak precision measurements~\cite{LEP-2}. Integrating out the $Z'$ in our model generates the effective four-fermion operators
\begin{equation}
\mathcal{L}_{\rm eff} \supset  
\frac{1}{2}
\left(\frac{g_2}{2 c_W}\right)^2
\left(\frac{g_V^e}{M_{Z'}}\right)^2
(\bar{e} \gamma_\mu e) (\bar{e} \gamma^\mu e),
\end{equation}
and
\begin{equation}
\mathcal{L}_{\rm eff} \supset  
\frac{1}{2}
\left(\frac{g_2}{2 c_W}\right)^2
\left(\frac{g_A^e}{M_{Z'}}\right)^2
(\bar{e} \gamma_\mu \gamma_5 e)(\bar{e} \gamma^\mu \gamma_5 e),
\end{equation}
which are constrained by measurements of $e^+ e^- \to e^+ e^-$ cross
sections at LEP.
The  upper limits
on the magnitudes of the electron-$Z'$ couplings are
\begin{equation}
|g_V^e| \le 
\frac{2c_W}{g_2}
\sqrt{4 \pi} \; \frac{M_{Z'}}{20.6 \mbox{ TeV}} 
\quad\text{and}\quad
|g_A^e| \le 
\frac{2c_W}{g_2}
\sqrt{4 \pi} \; \frac{M_{Z'}}{10.1 \mbox{ TeV}},
\end{equation}
which respectively hold in the case of vectorial and axial couplings. 

We scan the allowed range of couplings for two different   
masses of $Z'$, 
with the values  
$M_{Z'}=200$~GeV, 500~GeV
for scenarios involving muons, and 
$M_{Z'}=250$~GeV, 500~GeV
for those involving electrons. Our simulations were performed using \texttt{MadGraph\_aMC5@NLO}~\cite{Alwall:2014hca} and we validated our code in the special case of a sequential $Z'$  against the simulated cross section reported in Ref.~\cite{ATLAS:2016cyf} as a function of $M_{Z'}$. 

In our analyses we consider also the constraint from 
$B^0_s$--$\bar{B}^0_s$
oscillations in terms of the measured mass difference 
$\Delta M_s$
given by~\cite{Altmannshofer:2013foa}
\begin{equation}
	\label{eq:bosc}
\frac{\Delta M_s}{\Delta M_s^\mathrm{SM}}
\simeq 1 +
\frac{M_Z^2}{M_{Z'}^2}
\left[
(g^q_L)^2+(g^q_R)^2-9.7(g^q_L)(g^q_R)
\right]
\left(
\frac{g_2^2}{16\pi^2}
(V_{ts}^* V_{tb}^{\phantom{*}})^2 S_0
\right)^{-1}\,,
\end{equation}
where the contribution from the second term on the right-hand side is bounded by experiment to be in magnitude below the $10\%$ level. The SM loop function is $S_0 \simeq 2.3$.

As for the low-energy observations, including recent measurements by LHCb, we remark that a comprehensive statistical analysis of the Wilson coefficients would involve about 100 observables and a careful treatment of about 100 nuisance parameters. This is so technically challenging that instead approximate ``fast-fit'' techniques are favoured (see e.g.,~\cite{DAmico:2017mtc,Altmannshofer:2017yso}), although they require considerable computing resources. Since a comprehensive analysis is beyond the purpose of this paper, we adopt a pragmatic approach to understand the impact of the latest data. We perform an independent fit of the Wilson coefficients to LHCb measurements of $R_{K^*}$~\cite{LHCB} and $R_K$~\cite{Aaij:2014ora}, Belle measurements of $D^\prime_4$ and $D^\prime_5$~\cite{Wehle:2016yoi}, and LHCb measurements of $\mathcal{B}(B^0_s\to\mu^+\mu^-)$ and $\mathcal{B}(B^0\to\mu^+\mu^-)$~\cite{Aaij:2017vad}. 
Recent studies~\cite{DAmico:2017mtc,Geng:2017svp} performed global frequentist statistical analysis, whereas~\cite{Ciuchini:2017mik} performed a global Bayesian analysis with a subset of nuisance parameters and data with \texttt{HEPfit}~\cite{HEPfit}. Our methodology and its results are detailed in the next section.

\section{A first analysis} 
\label{sec:A first analysis}

We calculated the profile likelihood with \texttt{MultiNest}~\cite{Feroz:2008xx} interfaced with a modified version of \texttt{flavio}~\cite{Altmannshofer:2017fio}. Confidence intervals were found by Wilks' theorem (see~\cite{Fowlie:2016hew} for conventions and definitions). All nuisance parameters ($m_b$, $m_c$, CKM matrix elements and form factors) were fixed to their central values. Wilson coefficients were varied between $-5$ and $5$. We find reasonable agreement for best-fits, confidence intervals and significances with similar \texttt{flavio} results with Markov Chain Monte Carlo algorithms and fast-fit techniques for nuisance parameters~\cite{Altmannshofer:2017yso}. We estimate the significance of the preference for new physics versus the SM to be about $4.5\sigma$, in agreement with recent literature~\cite{DAmico:2017mtc,Geng:2017svp,Ciuchini:2017mik,Altmannshofer:2017yso,Capdevila:2017bsm}, but, as in the literature, there are important caveats about systematic uncertainties.   

Since it is a product of our \texttt{MultiNest} 
fit, we note that data favour the $C_9^\mu$ model by a Bayes factor of about $10^4$ versus the SM and about $10$ versus $C_{10}^\mu$, as well as versus $C_{9}^e$ or $C_{10}^e$. Since, however, we omitted important nuisance parameters that may alleviate tension in the SM, $10^4$ should be regarded as an upper bound to the Bayes factor versus the SM (and is, of course, sensitive to priors). 
A Bayes factor preference for 
$C_9^\mu$ versus electron Wilson coefficients was also noted in an earlier study of 
$R_K$~\cite{Ghosh:2014awa}.

\begin{figure}[t]
\centering
\includegraphics[scale=.4]{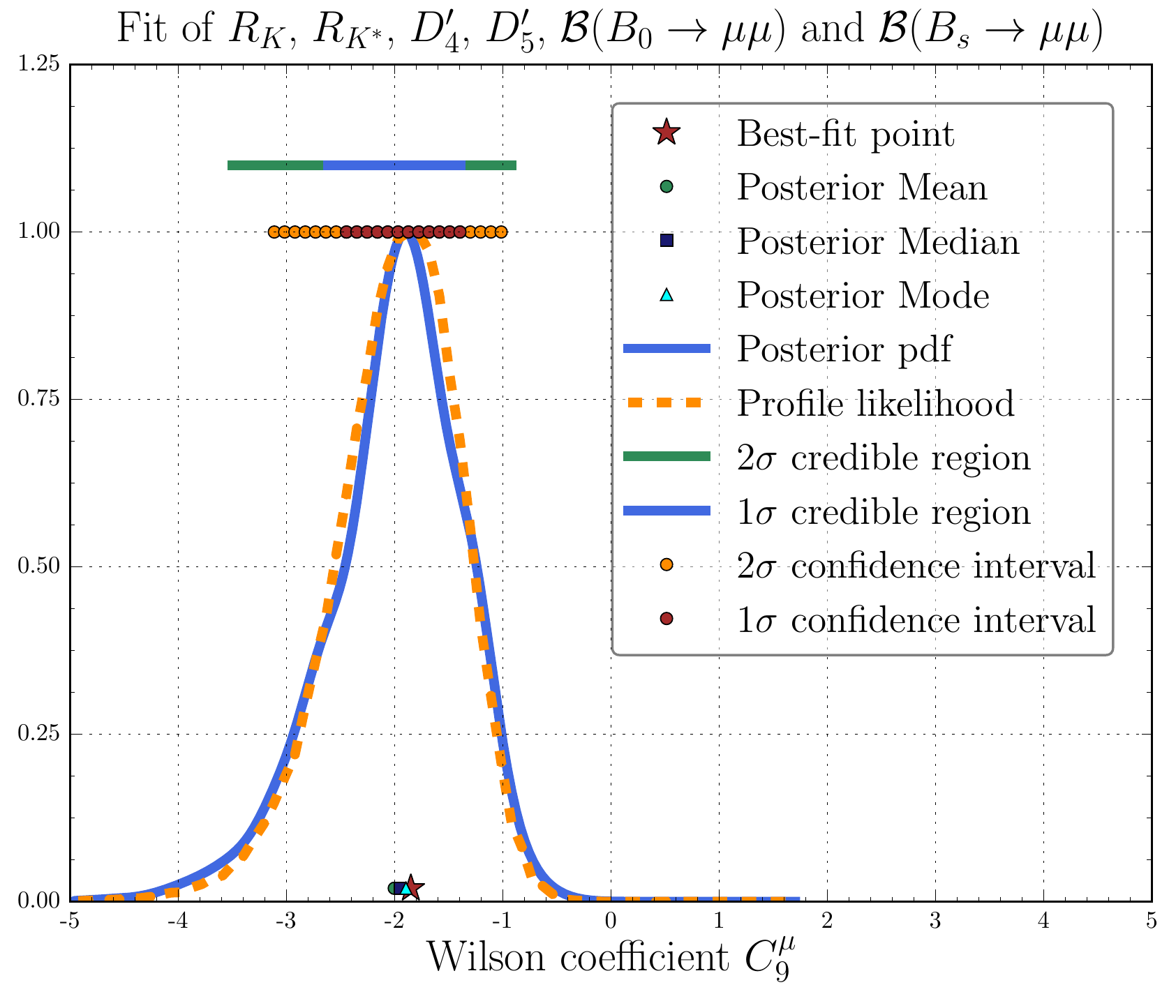}
\includegraphics[scale=.4]{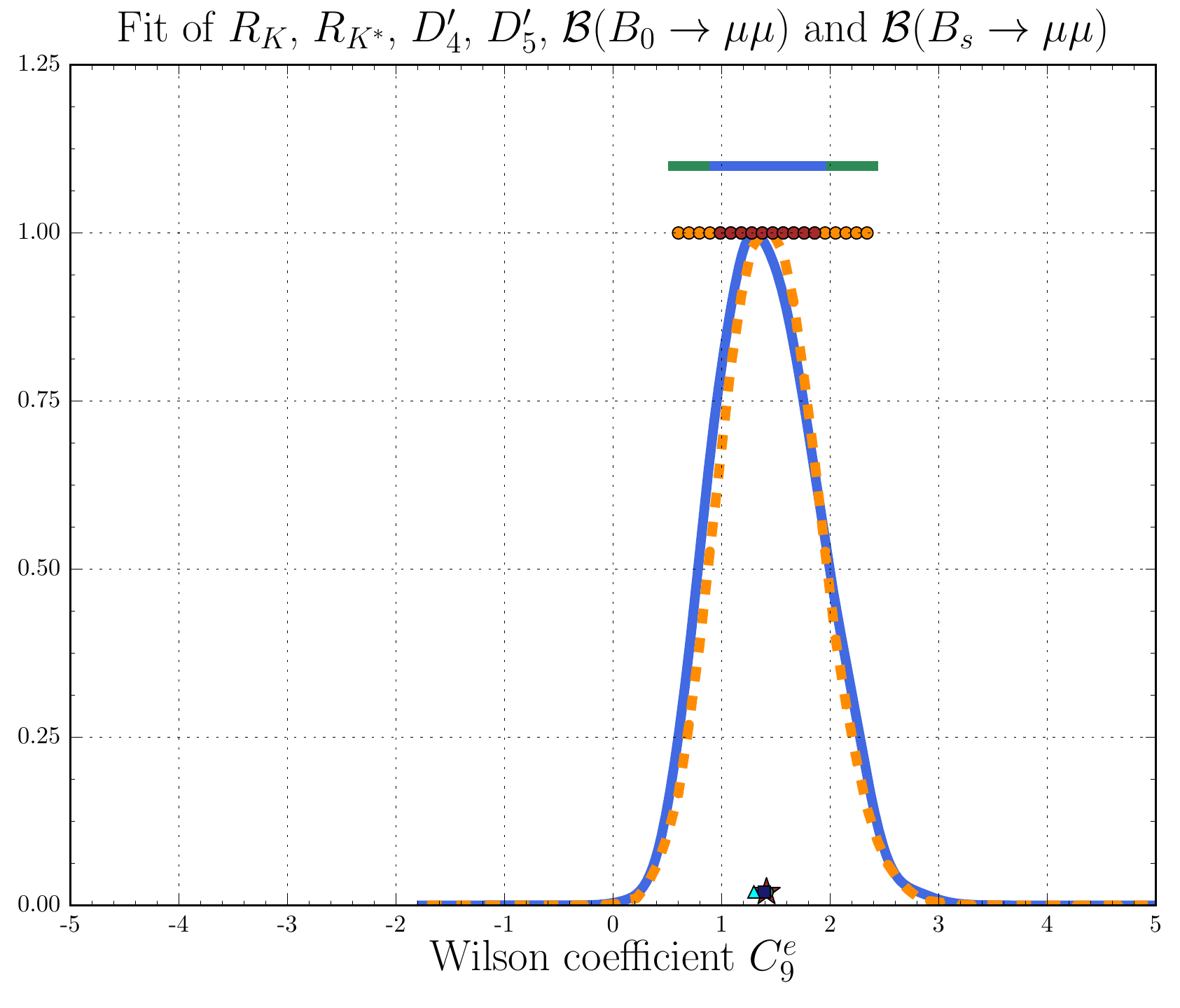}
\\
\includegraphics[scale=.4]{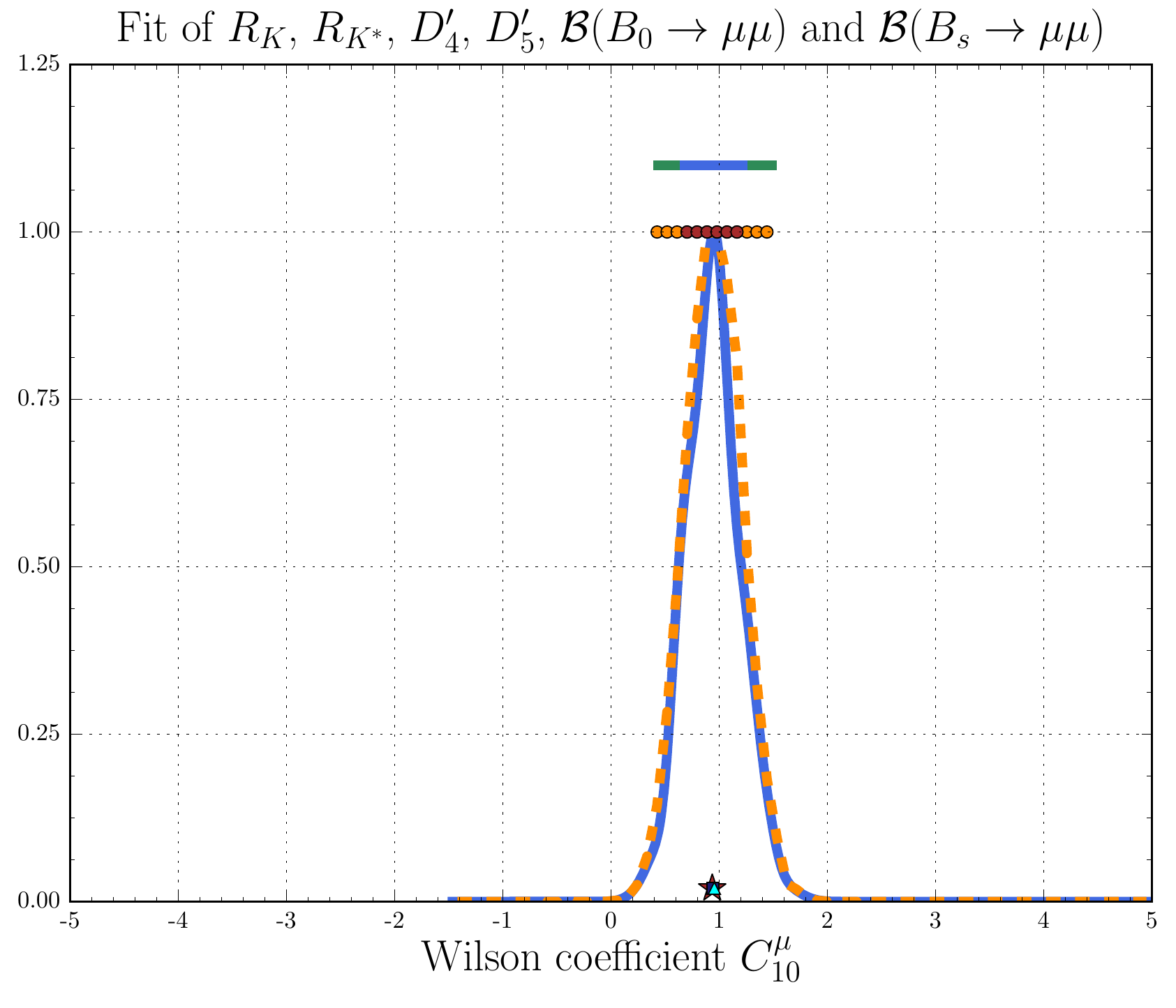}
\includegraphics[scale=.4]{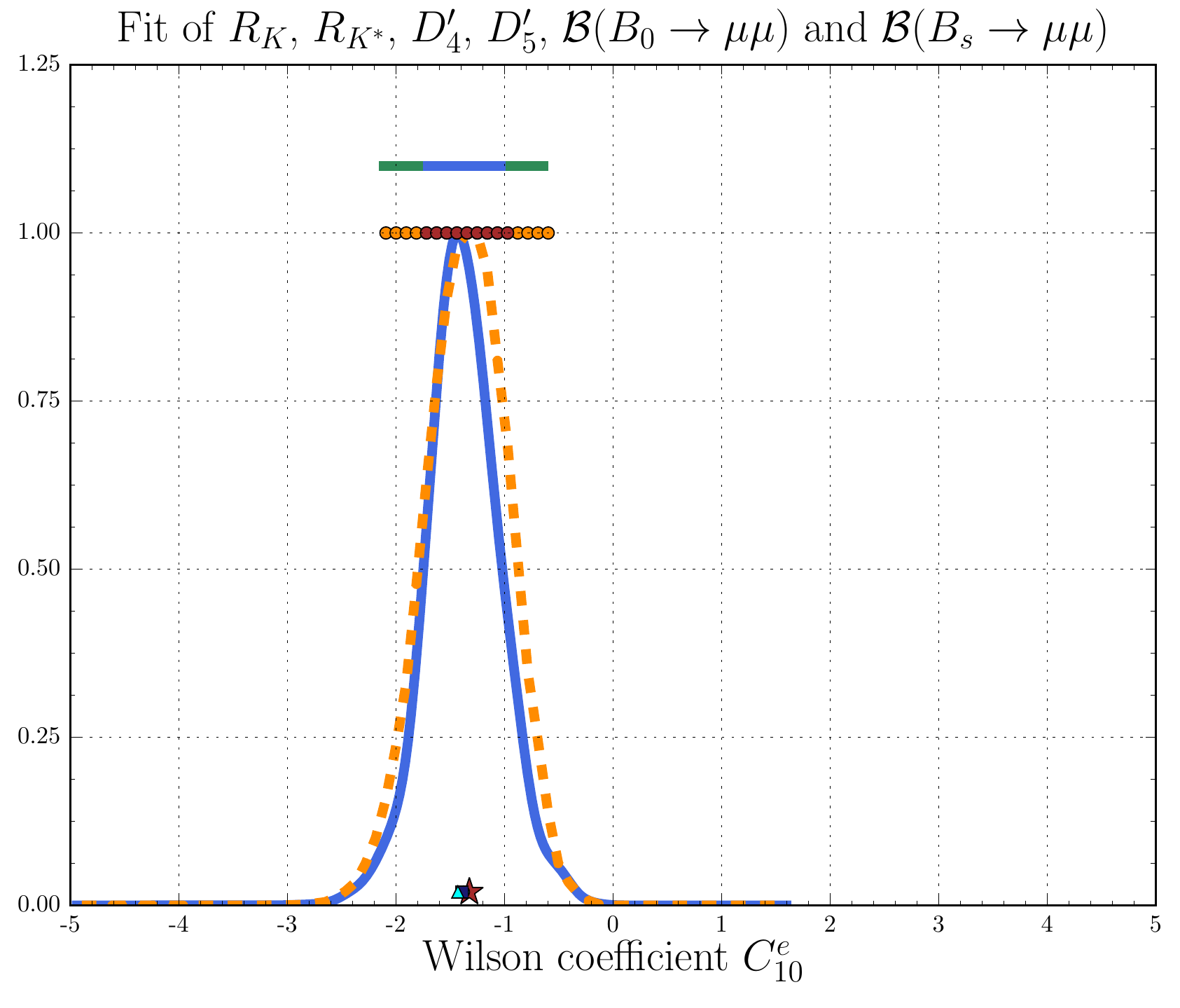}
\caption{Best-fits to $b\to s\ell^+\ell^-$ anomalies for the indicated Wilson coefficients.}
\label{fig:1}
\end{figure}

The results we obtained are presented in Figure~\ref{fig:1}, whereas Table~\ref{tab:Straub} proposes the results from Ref.~\cite{Altmannshofer:2017yso} for comparison. We find that the two analyses are in reasonable agreement given that the fits take into account different sets of observables.

We focus now on the constraints that collider and low energy experiments impose on the scenarios we delineated.

\newpage

\vfill

\begin{table}[b]
\centering
\begin{tabularx}{1\textwidth}{X<\centering X<\centering X<\centering X<\centering}
Coeff. & ~best fit~ & $1\sigma$ CL & $2\sigma$ CL\\
\hline
$C_9^{\mu}  $ & $-1.59$ & [$-2.15$, $-1.13$] & [$-2.90$, $-0.73$] \\
                    $C_{10}^{\mu}                   $ & $+1.23$ & [$+0.90$, $+1.60$] & [$+0.60$, $+2.04$] \\
$C_9^{e}                        $ & $+1.58$ & [$+1.17$, $+2.03$] & [$+0.79$, $+2.53$] \\
                    $C_{10}^{e}                     $ & $-1.30$ & [$-1.68$, $-0.95$] & [$-2.12$, $-0.64$] \\
\end{tabularx}
\caption{The values obtained for the Wilson coefficients from a fit to $b\to s\bar\ell\ell$ anomalies in Ref.~\cite{Altmannshofer:2017yso}.}
\label{tab:Straub} 
\end{table}

\clearpage
\newpage

\begin{figure}[t]
  \centering
    \includegraphics[width=.45\textwidth]{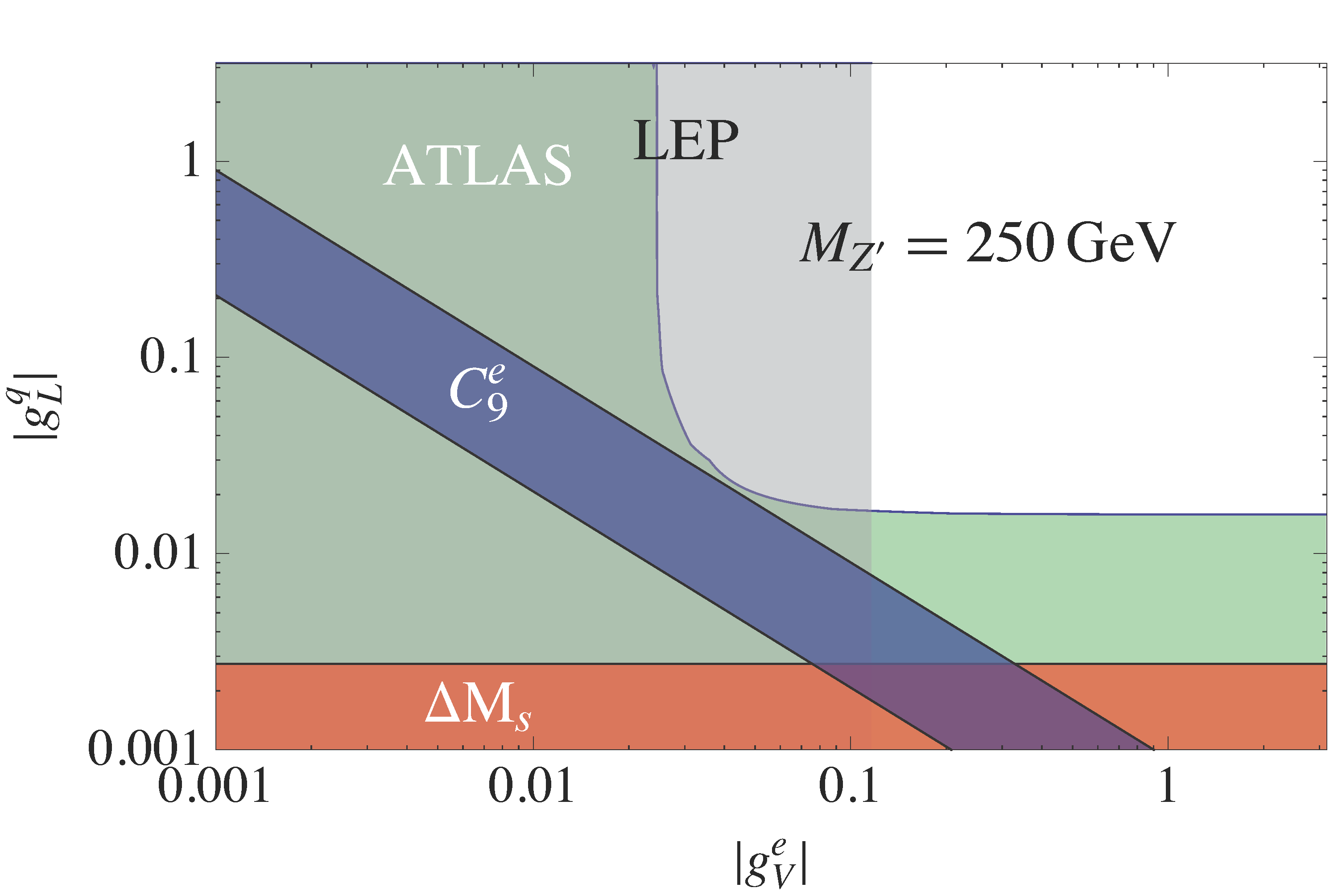}
	  \includegraphics[width=.45\textwidth]{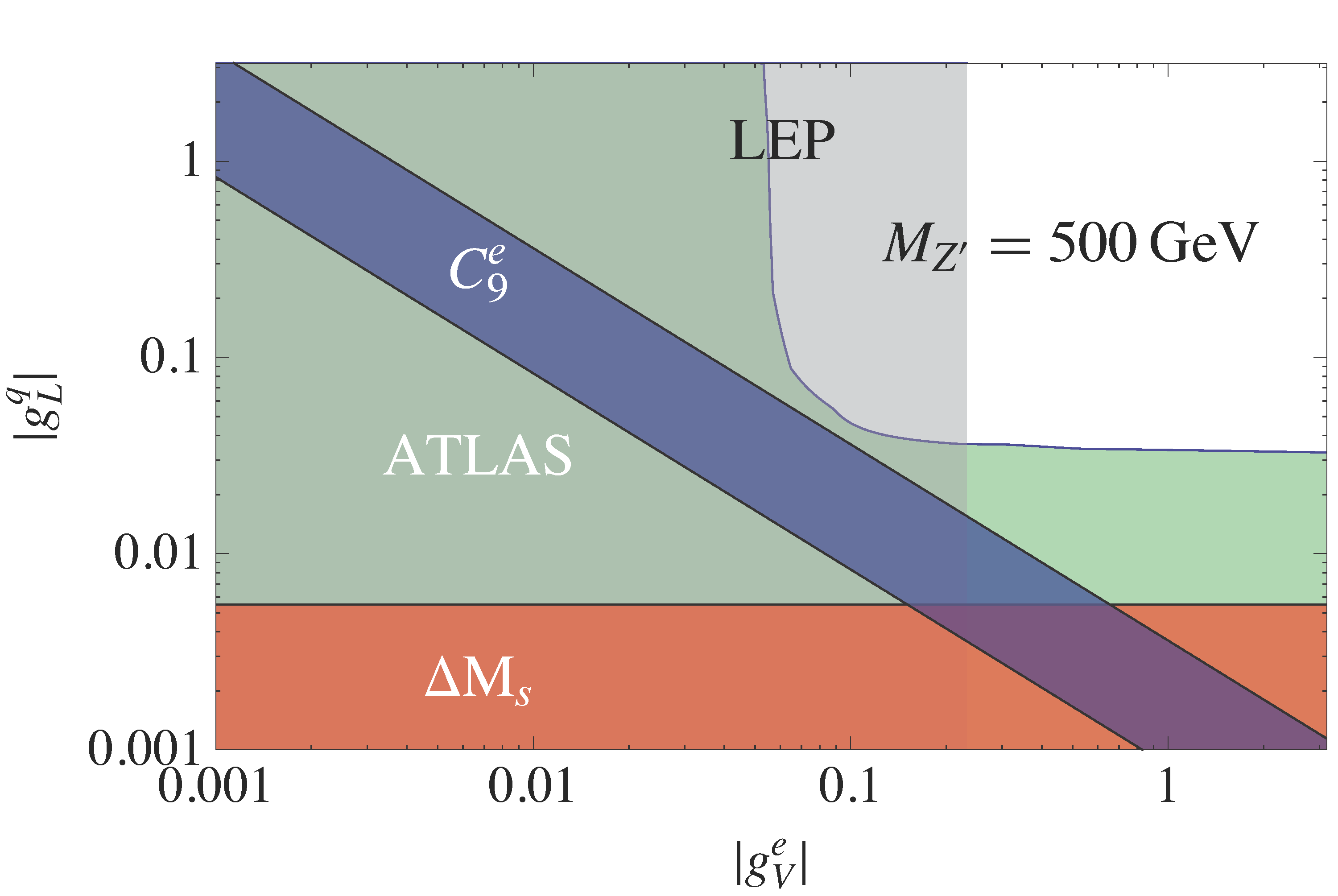}
	  \\
      \includegraphics[width=.45\textwidth]{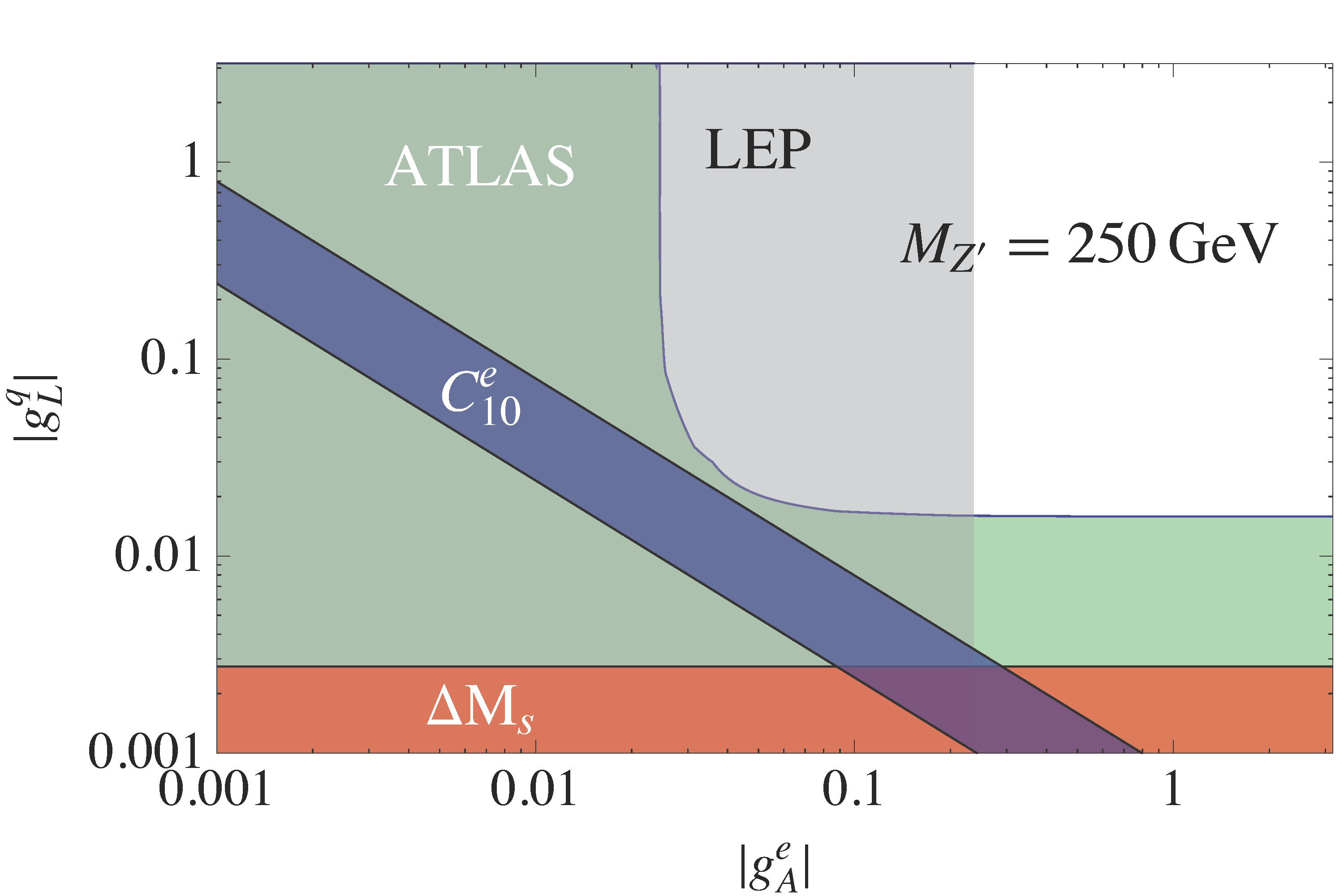}
  	  \includegraphics[width=.45\textwidth]{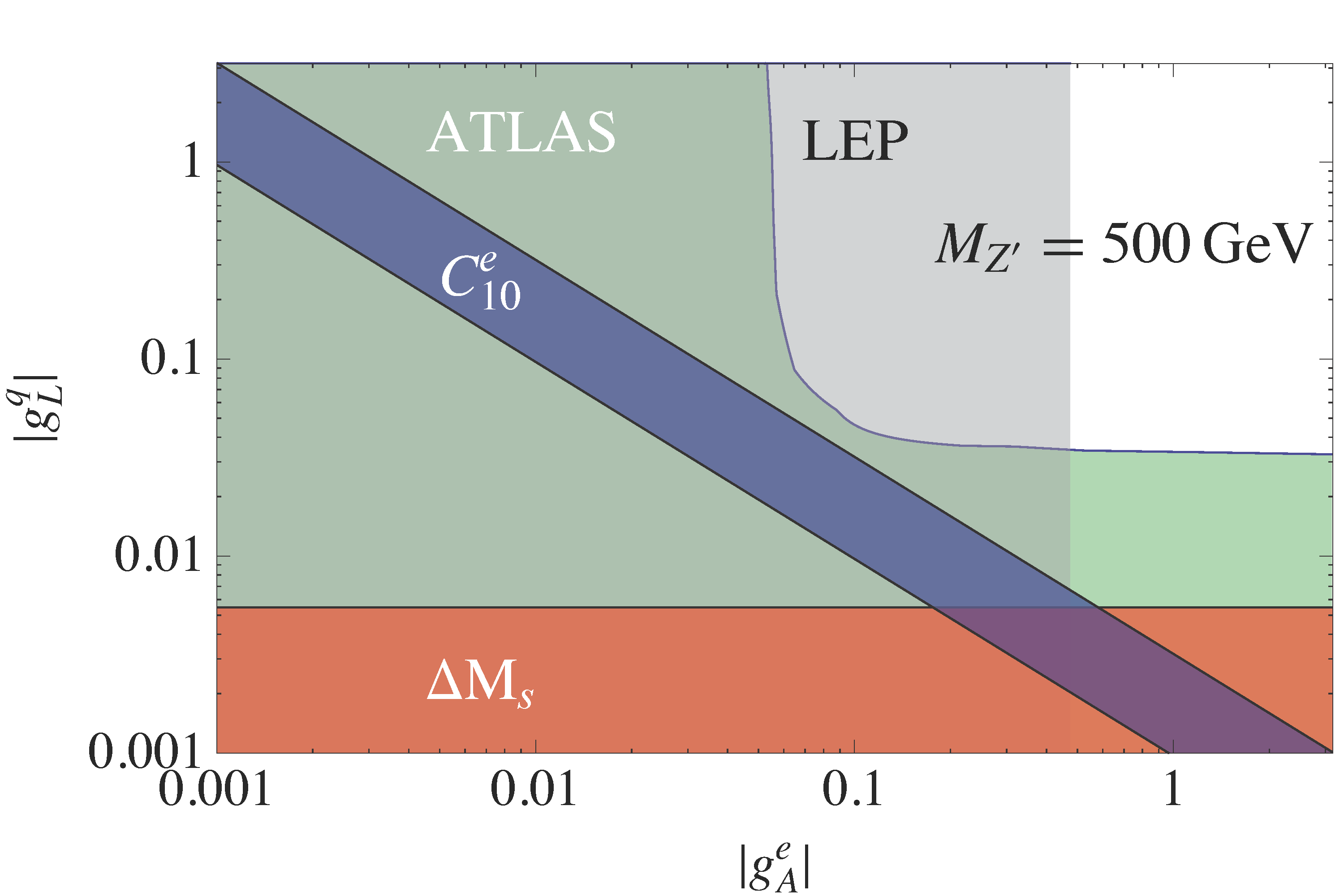}
  \caption{Experimental constraints on the scenario where the $Z'$ only couples to electrons. The shaded regions correspond to the allowed regions of parameter space. Upper panels: $C_9^e$ is non-zero and generated by a $Z'$ with mass 250 GeV (left) and 500 GeV (right). Lower panels: $C^e_{10}$ is non-zero and generated by a $Z'$ with mass 250 GeV (left) and 500 GeV (right).}
  \label{fig:e}
\end{figure}

\subsection{Scenarios with $C_9 ^e$  or $C_{10 }^e$ only} 
\label{sub:Scenario electron}

We present in Figure~\ref{fig:e} the results obtained for the scenario where new physics effects arise from the coupling of $Z'$ to electrons, and the relative effects are fully encapsulated in the Wilson coefficient $C_9^e$ (top panels), or $C_{10}^e$ (bottom panels).
The different shaded areas indicate allowed regions of 
parameter space after taking into account 
the various constraints.

The green area represents the region allowed by the ATLAS $Z'$ searches in the dielectron channel. The blue band represents the values of the coupling selected by the LHCb measurements of the indicated Wilson coefficient. We refer here to the values for the $2\sigma$ credible regions from our analysis presented in Figure~\ref{fig:1}. The red band illustrates instead the region of the parameter space allowed by the constraints on the mass difference $\Delta M_s$.  

\begin{figure}[t]
  \centering
    \includegraphics[width=.65\textwidth]{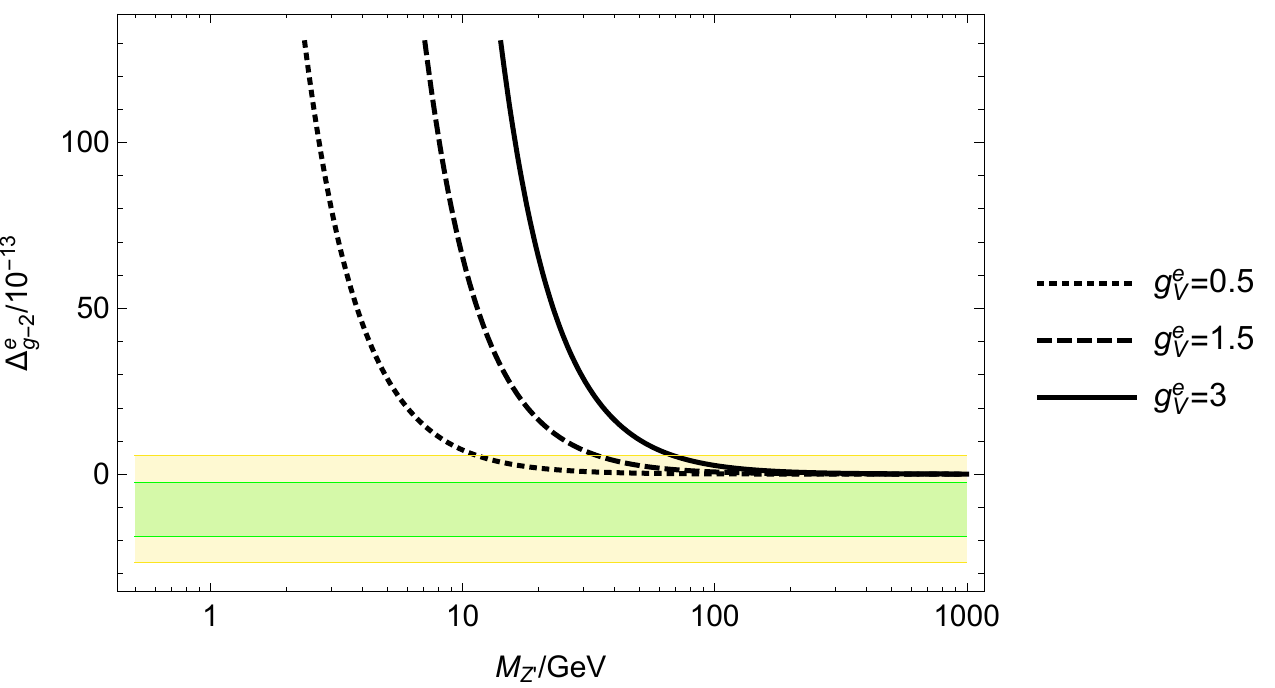}\\
	   \includegraphics[width=.65\textwidth]{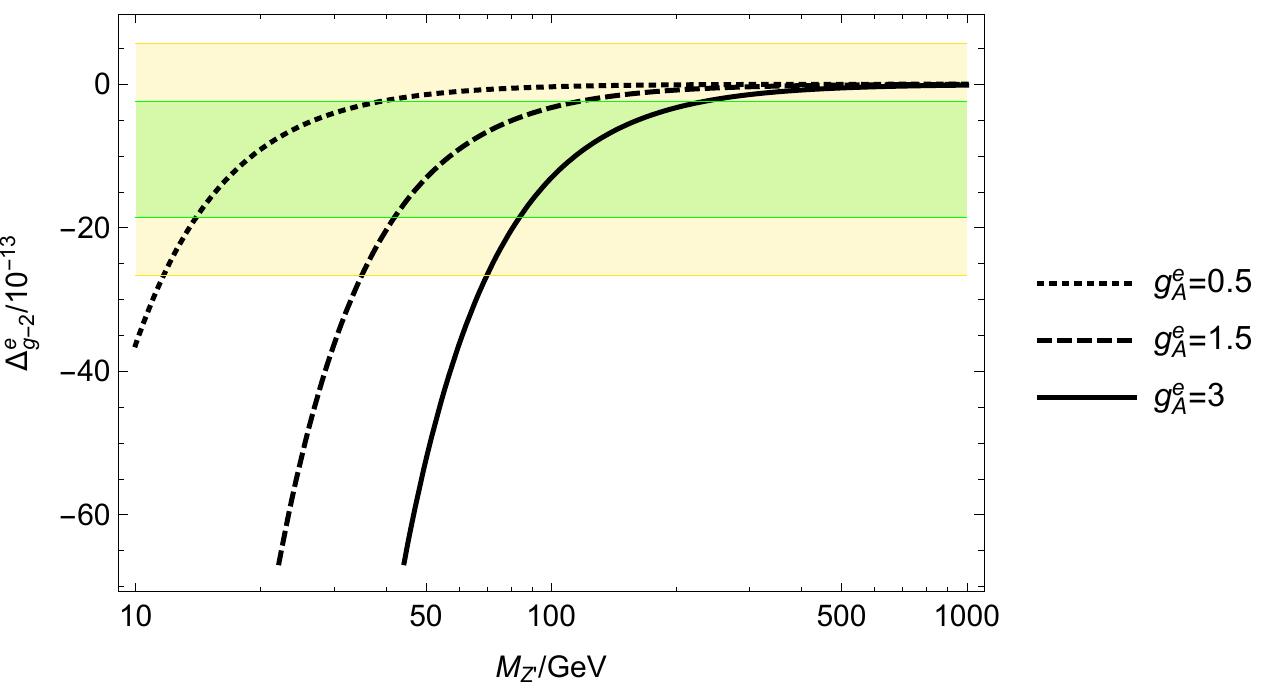}
  \caption{The $Z'$ contribution to the anomalous magnetic moment of the electron. The green and yellow bands represent the $1\sigma$ and $2\sigma$ confidence levels for the discrepancy between the measurement and the SM prediction. The top panel is for the scenario where all the new physics effects are contained in $C_9^e$, the bottom one for the case of $C_{10}^e$.}
  \label{fig:gm2e}
\end{figure}

As mentioned before, we also checked that the $Z'$ contribution to the electron $g-2$ does not spoil the current agreement between theory and experiment. A plot of the new contribution to this quantity is presented in Figure~\ref{fig:gm2e} as a function of $M_{Z'}$ for different values of the relevant coupling. The results in the top panel holds for the scenario where all the new physics effects are contained in $C_9^e$, the bottom one for the case of $C_{10}^e$. We can see that values of $M_{Z'}\lesssim 50 $ GeV would negatively impact on the prediction for the anomalous magnetic moment of the electron.

We remark that these scenarios where the $Z'$ only couples to electrons, besides quarks, are strongly constrained by the bounds from LEP-II~\cite{LEP-2}. The gray shade in Figure~\ref{fig:e} denotes the area of the parameter space that evades the latter. We can see that the scenario with non-zero $C_9^e$ generated from a vector coupling is more tightly constrained than $C_{10}^e$ by LEP electroweak precision measurements and $B^0_s$--$\bar{B}^0_s$ oscillations. 
Indeed in the case of non-zero $C_{10}^e$ a larger region of the parameter space is still allowed by all the considered experimental limits. A $Z'$ boson with axial vector couplings would also induce a negative contribution to the $g-2$ of the electron, 
and is therefore amenable to reducing the tension between the measurements and the SM prediction. Although in the simple models we are considering this contribution is not substantial, we remark that the preference for $C_{10}^e$ based on $g-2$ is consistent with the picture emerging from the other constraints analysed.

\FloatBarrier

\subsection{Scenarios with $C_9^\mu$ and $C_{10}^\mu$ only} 
\label{sub:Scenario II}

\begin{figure}[t]
  \centering
      \includegraphics[width=.45\textwidth]{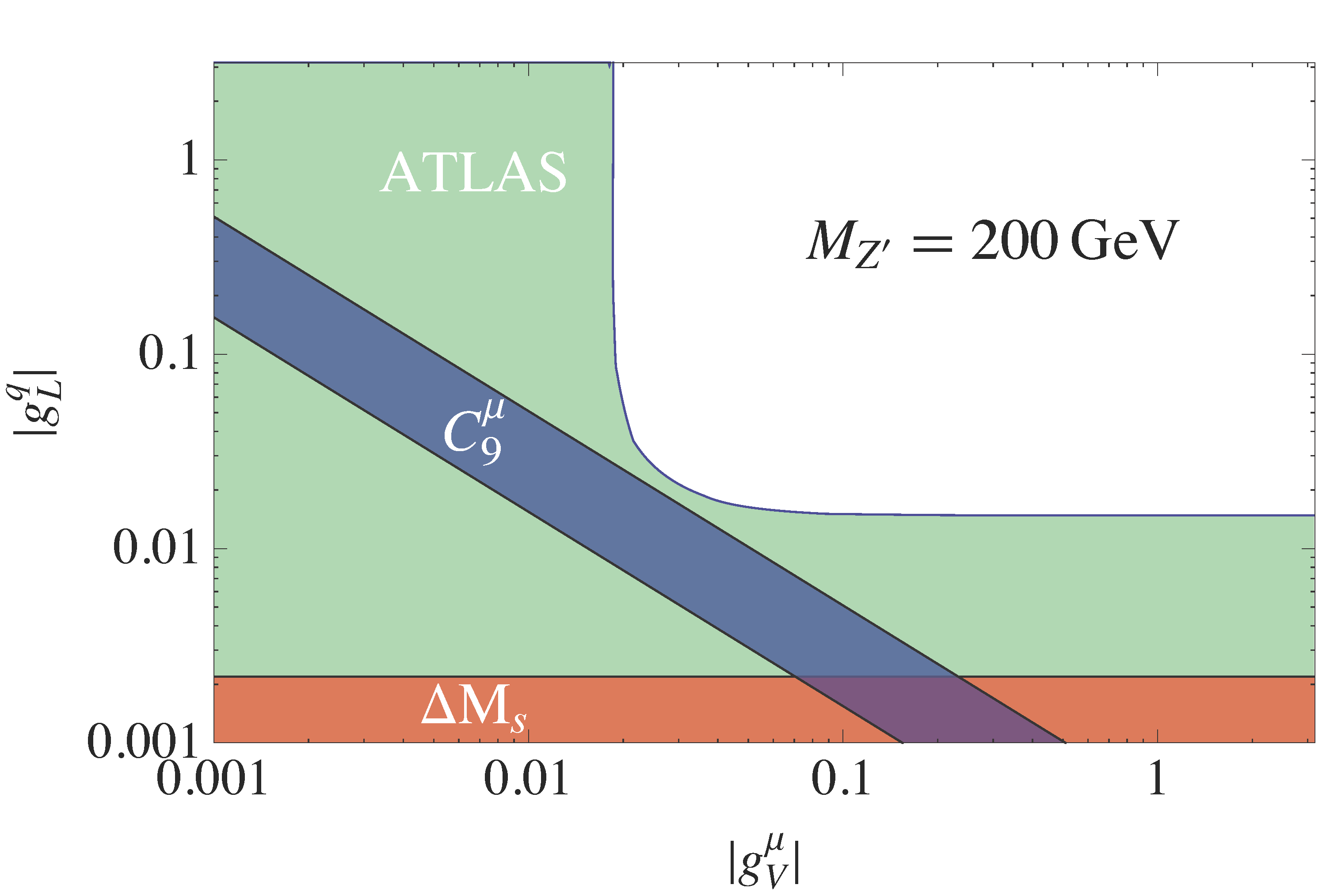}    
	  \includegraphics[width=.45\textwidth]{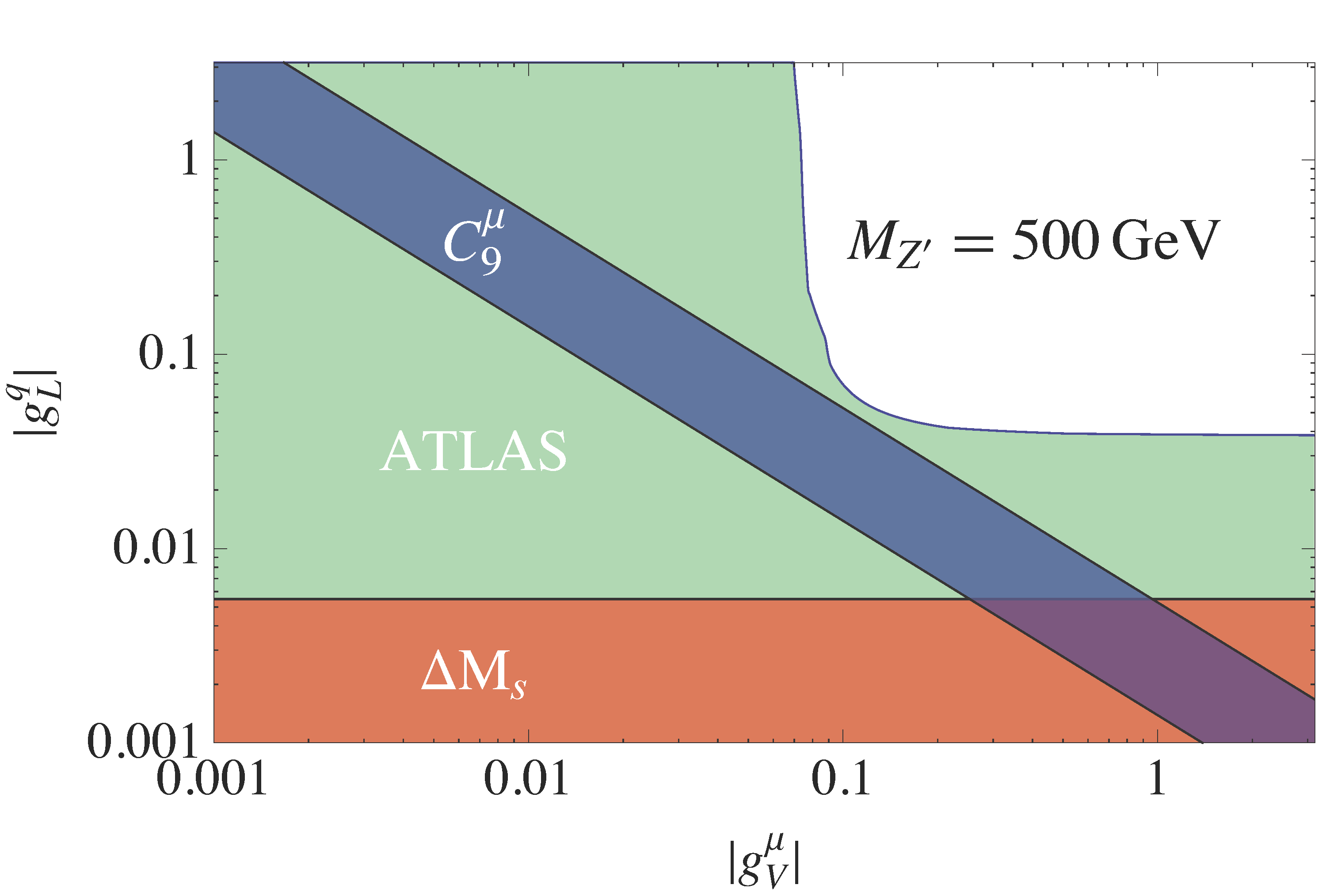}
	  \\
      \includegraphics[width=.45\textwidth]{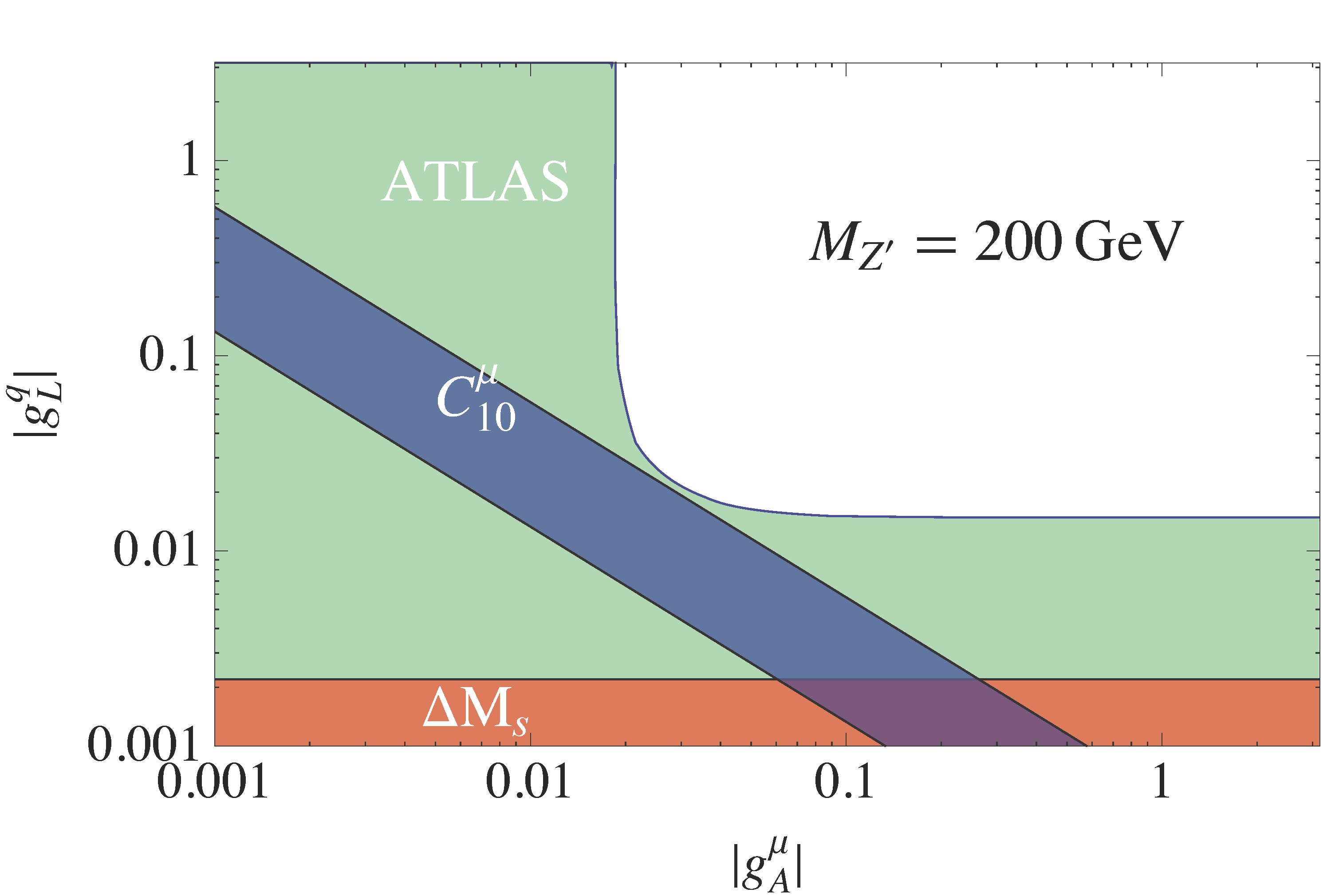}
  	  \includegraphics[width=.45\textwidth]{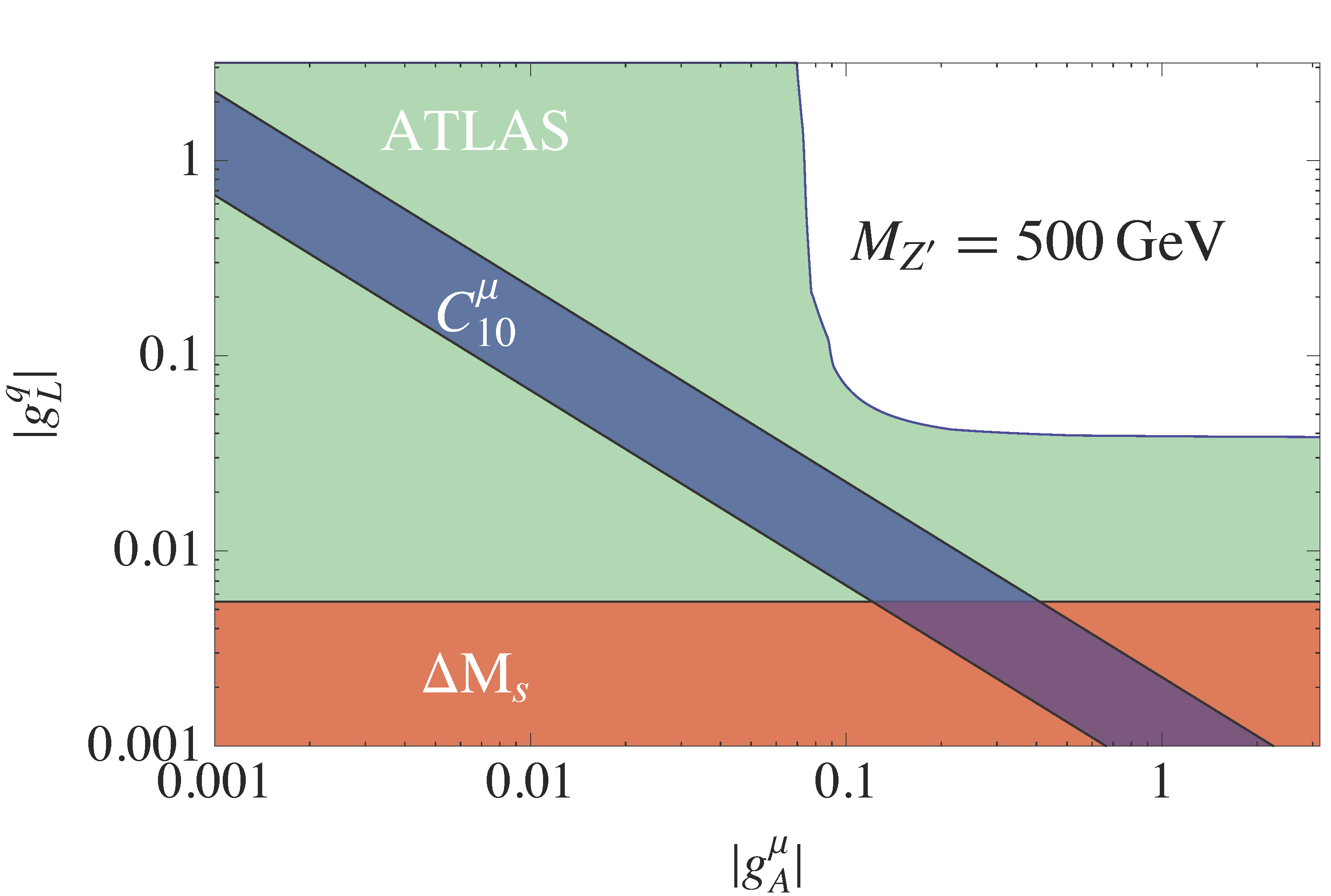}
  \caption{Experimental constraints on the scenario where the $Z'$ only couples to muons. The shaded regions correspond to the allowed regions of parameter space. Upper panels: $C_9^\mu$ is non-zero and generated by a $Z'$ with mass 200 GeV (left) and 500 GeV (right). Lower panels: $C^\mu_{10}$ is non-zero and generated by a $Z'$ with mass 200 GeV (left) and 500 GeV (right).}
  \label{fig:muons}
\end{figure}

We present in Figure~\ref{fig:muons} the analogous results obtained for the case in which, 
besides quarks, the speculated $Z'$ couples exclusively to muons.  

Again we show in the top panel the case where new physics affects $C_9^\mu$ only, and the bottom panels cover the complementary case of $C_{10}^\mu$. As before, the green area represents the region of the parameter space allowed by the ATLAS $Z'$ searches, here for the dimuon channel. The blue band represents the values of the coupling selected by the LHCb measurements of the indicated Wilson coefficient, according to the $2\sigma$ credible intervals shown in Figure~\ref{fig:1}. The red band is the region of the parameter space allowed by the constraints on the mass difference $\Delta M_s$. We remark that LEP measurements do not constrain the couplings in these scenarios. 

\begin{figure}[t]
  \centering
    \includegraphics[width=.65\textwidth]{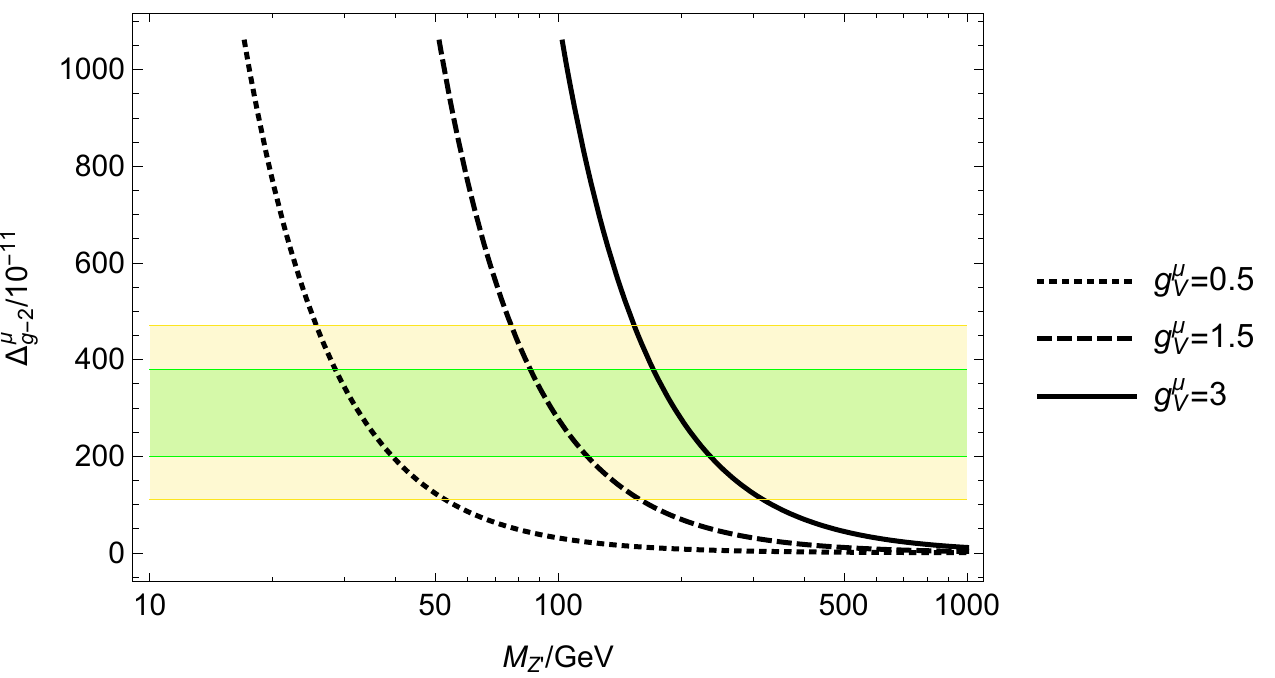}
  \caption{The $Z'$ contribution to the anomalous magnetic moment of the muon. The green and yellow bands represent the $1\sigma$ and $2\sigma$ confidence levels for the discrepancy
between the measurement and the SM prediction. }
  \label{fig:gm2m}    
\end{figure}

The $Z'$ contribution in the $C_9^\mu$ scenario can help to bridge the discrepancy
between the 
measured value of the muon magnetic moment and the SM prediction. 
As we can see in 
Figure~\ref{fig:gm2m}, the $Z'$ contribution is potentially able to reduce such a discrepancy below the $1\sigma$ level for moderate values of the $Z'$-muon couplings for $Z'$ masses up to about $250$ GeV.
The solid black line in this region with
$g^\mu_V = 3$ and $M_{Z'}\simeq 200$~GeV
is compatibile with the parameter space
selected by the LHCb data.
In this case, the physical coupling 
$g^\mu_V (g_2/2c_W) \simeq 0.3 \sqrt{4\pi}$
is large, but marginally within the perturbative regime.
The corresponding region in Figure~\ref{fig:muons}
is allowed, 
when the top left panel is extrapolated to
$g^q_L \simeq 10^{-4}$.

As for the $C_{10}^\mu$ case (not shown), the presence of a $Z'$ in the low-mass range can only worsen the theory prediction as the relative contribution, of negative sign, further lowers the value of the latter. 

\newpage

\section{Conclusions}
\label{sec:Conclusions}

In this paper we considered four different new physics scenarios which could explain the anomalous $B$-physics results from LHCb and argued that measurements of the anomalous magnetic moments of muon and electron could break the degeneracy in the $C^{\mu(e)}_9$--$C^{\mu(e)}_{10}$ 
Wilson coefficients.

We find that a $Z'$-boson which couples only to electrons can produce the correct values for $R_K$ and $R_{K^*}$, but such a scenario is 
strongly constrained by the precision measurements from LEP. However,
if the $Z'$ coupling to $e^+e^-$ is purely axial, 
the allowed region of parameter space is larger.
Such an axially coupled $Z'$ also generates a negative contribution to the anomalous magnetic moment of the electron and could therefore potentially accommodate both anomalies.

We also considered scenarios where the $Z'$-boson couples only to muons. In this case, a vectorial coupling to the $Z'$ is favoured by the Bayesian approach employed for the performed fit. Such a scenario produces also a positive contribution to the $g-2$ of the muon and therefore could help to alleviate the tension between the $g-2$ measurement and the SM prediction. This scenario can also explain the LHCb anomalies while avoiding all the remaining constraints from high energy searches, provided the vectorial coupling to muons is large ($g_V^\mu > 0.1$). An axial coupling of the $Z'$ to muons can also explain the LHCb results, but increases the disagreement of the $g-2$ with measurements.

\acknowledgments

We thank Christian Veelken for useful discussions. 
This work was supported by the EU through the ERDF CoE program grant TK133.
The work of CM is supported by the 
``Angelo Della Riccia'' foundation.

\bibliographystyle{JHEP}
\bibliography{references}

\end{document}